\documentclass[12pt]{article}
\usepackage[utf8]{inputenc}
\usepackage{amsmath, amsfonts}
\usepackage{graphicx}
\usepackage[margin=2cm]{geometry}
\usepackage{subcaption}
\usepackage{float}
\usepackage{xcolor}
\usepackage[hidelinks]{hyperref}\usepackage{ulem}

\title{Numerical Issues for a Non-autonomous Logistic Model}
\author{Marina Mancuso$^{a,b,*}$, Carrie A. Manore$^{c}$, Kaitlyn Martinez$^{a}$, Fabio Milner$^{b}$}

\begin{document}

\maketitle

\noindent $a$: A-1 Information Systems and Modeling, Los Alamos National Laboratory, Los
Alamos NM, 87544 USA

\noindent $b$: School of Mathematical and Statistical Sciences, Arizona State University, Tempe AZ, 85287 USA

\noindent $c$: T-6 Theoretical Biology and Biophysics, Los Alamos National Laboratory, Los
Alamos NM, 87544 USA \newline

\noindent \textit{*Corresponding author: mmancuso@lanl.gov}

\begin{abstract}
The logistic equation has been extensively used to model biological phenomena across a variety of disciplines and has provided valuable insight into how our universe operates. Incorporating time-dependent parameters into the logistic equation allows the modeling of more complex behavior than its autonomous analog, such as a tumor’s varying growth rate under treatment, or the expansion of bacterial colonies under varying resource conditions. Some of the most commonly used numerical solvers produce vastly different approximations for a non-autonomous logistic model with a periodically-varying growth rate changing signum. Incorrect, inconsistent, or even unstable approximate solutions for this non-autonomous problem can occur from some of the most frequently used numerical methods, including the lsoda, implicit backwards difference, and Runge-Kutta methods, all of which employ a black-box framework. Meanwhile, a simple, manually-programmed Runge-Kutta method is robust enough to accurately capture the analytical solution for biologically reasonable parameters and consistently produce reliable simulations. Consistency and reliability of numerical methods are fundamental for simulating non-autonomous differential equations and dynamical systems, particularly when applications are physically or biologically informed.
\end{abstract}

\section{Introduction}

Density-dependent biological phenomena such as tumor growth, fishery management, and mosquito populations can all be modeled with logistic growth dynamics \cite{KuangNagyEikenberry2016, NesslageWilberg2012, Rubel2008}. The classical logistic growth model was first conceptualized by Fran\c cois Verhulst and has two parameters-- the net growth rate and the carrying capacity \cite{Verhulst1838}. The net growth rate is the difference between the recruitment and removal rates of the population, and the carrying capacity is the upper threshold value of the population that can be biologically sustained. Under logistic growth, a population's rate of change decreases linearly as its size approaches the carrying capacity. Most applications using logistic growth dynamics assume the biological parameters remain constant with respect to time. However, one such extension of the logistic model is the non-autonomous version, which considers a time-dependent net growth rate and/or carrying capacity. 

Various theoretical aspects of non-autonomous logistic growth have been explored, including establishing the existence and uniqueness of a general solution \cite{VanceCoddington1989} and stability of periodic solutions \cite{JiangShi2005}. Additionally, Banks provides several applications of non-autonomous logistic models ranging from agricultural populations to railroad mileage \cite{Banks1994}. In all aforementioned cases, however, the infimum of the time-varying net growth rate is assumed to be positive. Nevertheless, it may be more appropriate for some applications to allow for the possibility of a negative net growth rate to capture more general behavior. For example, tumor cells may die faster than they grow under treatment, or mosquito populations can experience negative net growth rates under seasonal temperature variation. However, employing a negative net growth rate creates additional modeling challenges by modifying the logistic growth's dynamic behavior. Rather than the carrying capacity acting as a maximum value representing a saturated population, this parameter becomes the minimum threshold necessary to sustain population growth under a negative net growth rate \cite{BoyceDiPrima2001}. Numerical methods must retain behavioral characteristics of the systems they represent in order to be of value. Numerical methods for non-autonomous logistic growth should ensure that the simulated population remains nonnegative and bounded.

Built-in numerical solvers in popular simulation software such as Matlab or Python for a case of the non-autonomous logistic model are not robust enough to capture the behavior of the analytical solution, and can produce biologically invalid or unstable results. While the continuous logistic model allows for unique solutions for initial value problems, numerically solving the ordinary differential equation makes it inherently discrete. As a result, numerical simulations may produce chaotic behavior as observed in the discrete logistic equation \cite{BoyceDiPrima2001, Teschl2012}. Moreover, a simple, manually-programmed numerical solver avoids these aforementioned issues and outperforms the built-in numerical solvers available. The black box nature of built-in numerical solvers creates difficulties in pinpointing numerical inconsistencies of simulations.

The motivating example of the non-autonomous logistic model takes the following form:

\begin{subequations}
\renewcommand{\theequation}{\theparentequation.\arabic{equation}}
\label{eq: model}
\begin{equation}
P'(t) = r(t) P(t) \left(1 - \frac{P(t)}{K} \right),
\end{equation}
\noindent where $P(t)$ is the population size at time $t$, $K$ is the carrying capacity, and $r(t)$ is the time-varying intrinsic net growth rate with units of day$^{-1}$. A periodic function with a 365 day period models the time-varying growth rate:  
\begin{equation}
    r(t) = r_b - r_s \cos\left(\frac{2 \pi t}{365} \right),  
\end{equation}
\end{subequations}

\noindent where $r_b$ is the baseline, or mean net growth rate, and $r_s$ is the amplitude scaling factor. No restrictions are employed on $r_b$ or $r_s$, but most examples shown here have $|r_b| < |r_s|$ to allow $r(t)$ to change sign. A realistic application for Eq.~(\ref{eq: model}) may be to represent mosquito populations in a temperate climate, where temperature and precipitation are time-varying factors affecting mosquito growth in a nonlinear way \cite{ShamanDay2007}.

When supplied with initial condition $P(0) = P_0$, Eq.~(\ref{eq: model}) becomes an initial value problem with the explicit analytical solution,

\begin{subequations}
\renewcommand{\theequation}{\theparentequation.\arabic{equation}}
\label{eq: true solution}
\begin{equation}
    P(t) = \frac{P_0 K}{(K - P_0) \exp^{-f(t)} + P_0},
\end{equation}
with,
\begin{equation}
    f(t) = r_b t - \left(\frac{365 r_s}{2 \pi} \right) \sin \left( \frac{2 \pi t}{365} \right).
\end{equation}
\end{subequations}

\section{Simulations of numerical solvers}

Six numerical solvers implemented in Python compare approximations of Eq.~(\ref{eq: model}) to its analytical solution using various $r_b$ and $r_s$ values:

\begin{enumerate}
    \item odeint: lsoda method \cite{Hindmarsh1983}, implemented with SciPy's \texttt{odeint} function \cite{odeint}
    \item LSODA: lsoda method, implemented with SciPy's \texttt{solve\_ivp} function \cite{solve_ivp}
    \item BDF: implicit backwards differentiation, implemented with SciPy's \texttt{solve\_ivp} function \cite{solve_ivp}
    \item RK23: second-order Runge-Kutta method, implemented with SciPy's \texttt{solve\_ivp} function \cite{solve_ivp}
    \item RK45: forth-order Runge-Kutta method, implemented with SciPy's \texttt{solve\_ivp} function \cite{solve_ivp}
    \item manual RK4: fourth-order Runge-Kutta method, manually-programmed \cite{BoyceDiPrima2001, Teschl2012}
\end{enumerate}

The odeint and LSODA methods are flexible for stiff problems, and the first five tests employ adaptive or quasi-adaptive step size algorithms \cite{odeint, solve_ivp}. The manual RK4 test uses a fixed step size of $\Delta t = 1$ day. The odeint, BDF, RK45, and manual RK4 methods are used to simulate a more complex non-autonomous logistic growth model, where both net growth rate and carrying capacity vary periodically with time. This more complex model does not have an analytical solution for comparison. 
\subsection{Periodic growth rate with analytical solution}

The numerical approximations from the six tests using four Parameter Sets (PS1--PS4) are compared to their analytical solutions in Table (\ref{tab: RMSE}). The manual RK4 method shows the closest approximation to the analytical solution across all parameter sets. Furthermore, the results of simulations using the manual RK4 method are very similar, and indeed more accurate, when using decreasing step size $\Delta t < 1$, until round-off errors accumulate and halt the improvement in accuracy.

\begin{table}[h]\centering
\caption{Root mean squared error (RMSE) values from numerical simulations of non-autonomous logistic model Eq.~(\ref{eq: model}) for four Parameter Sets (PS1--PS4) with carrying capacity $K = 200,000$ and initial condition $P(0) = 1$. A dash '---' indicates that the simulation produced unbounded output and the root mean squared error could not be determined. Bold values show the lowest RMSE for the parameter set.}
\label{tab: RMSE}
\begin{tabular}{lcccccccc}
   & $r_b$ & $r_s$ & odeint   & LSODA    & BDF    & RK45   & RK23   & Manual RK4     \\\hline
PS1 & 0.05 & 0.15 & 14 & 497   & 913 & 552 & -----    & \textbf{5.62}  \\
PS2 & -0.002 & 0.25 & 517   & 779   & 147 & 671 & -----    & \textbf{18.95} \\
PS3 & 0.01 & 0.25 & 1,355 & 6,765 & -----    & -----    & -----    & \textbf{23.89} \\
PS4 & 0.005 & 0.05 & 3.53     & 765   & 505 & 757 & 663 & \textbf{1.05} 
\end{tabular}
\end{table}

\begin{figure}\centering
    \includegraphics[width=0.8\linewidth]{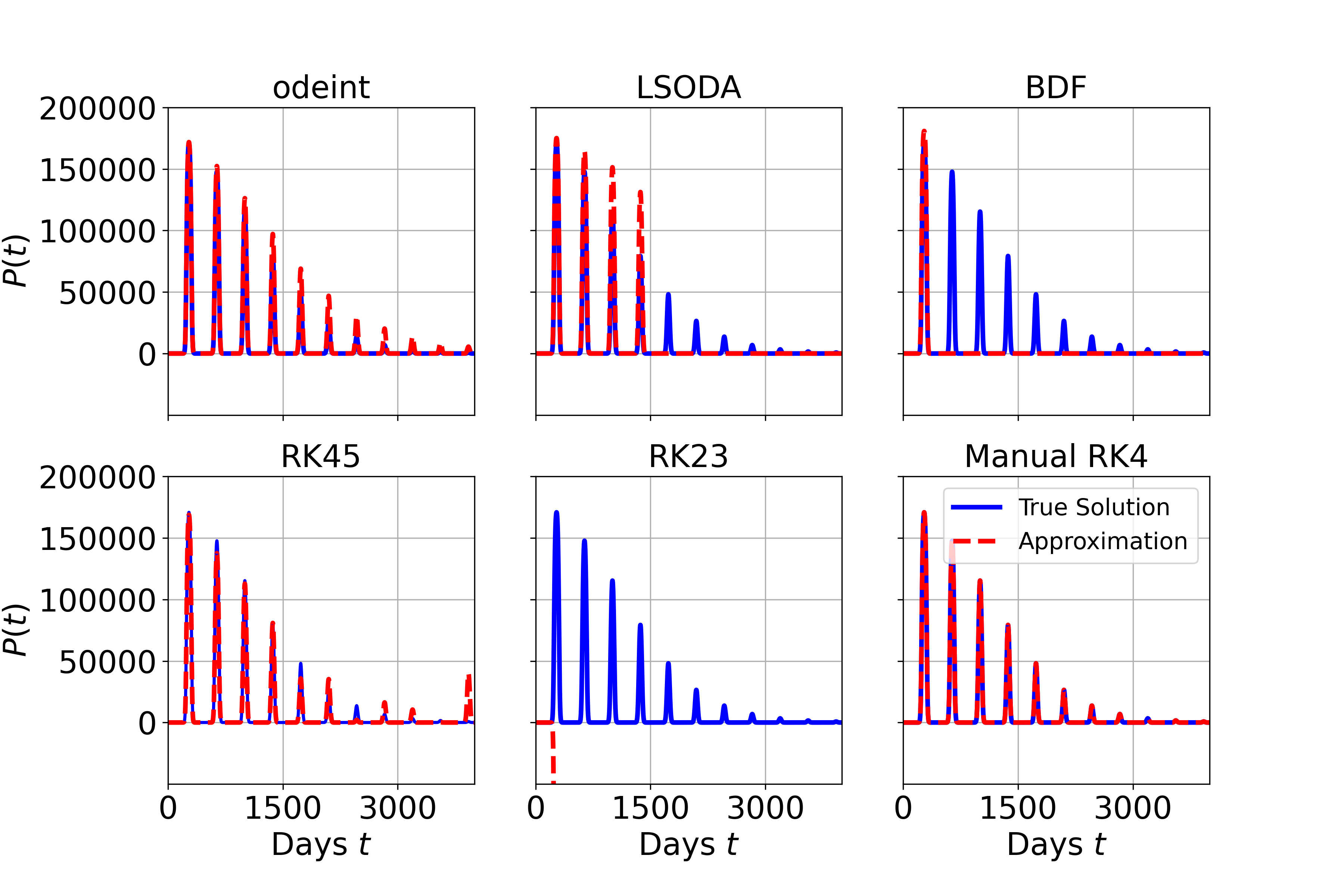}
    \caption{Numerical approximations (red dashed curves) of Eq.~(\ref{eq: model}) and analytical solutions (blue curves) using $r_b = -0.002$ and $r_s = 0.25$. Simulations have $K = 200,000$ with initial condition $P(0) = 1$.}
    \label{fig: PS2}
\end{figure}

\begin{figure}\centering
  \includegraphics[width=0.8\linewidth]{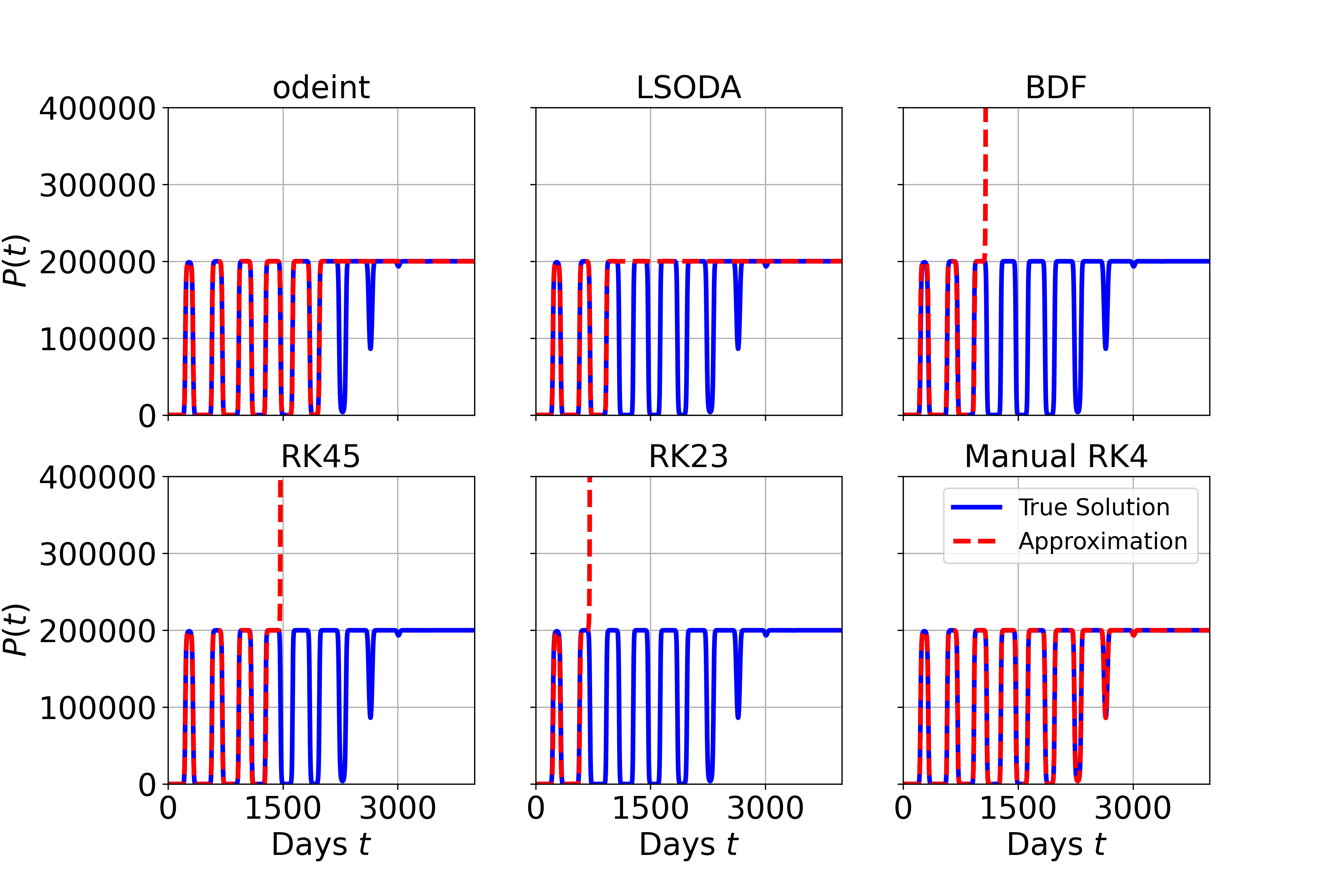}
  \caption{Numerical approximations (red dashed curves) of Eq.~(\ref{eq: model}) and analytical solutions (blue curves) using $r_b = 0.01$ and $r_s = 0.25$. Simulations have $K = 200,000$ with initial condition $P(0) = 1$. To enhance visual comparison, figures for the unstable simulations only show up to $P(t) = 400,000$.}
  \label{fig: PS3}
\end{figure}

Simulations for PS2 and PS3 from the six tests are shown in Figure \ref{fig: PS2} and Figure \ref{fig: PS3}, respectively. PS2 is an example of simulations decaying to the zero equilibrium, while PS3 shows simulations approaching the carrying capacity equilibrium. Each numerical solver shows significantly different qualitative behavior in both parameter sets. The manual RK4 method was the only test to produce behavior consistent with the analytical solution for both parameter sets. The odeint method overshoots the magnitude of the oscillations for PS2, and falls short of capturing all oscillations for PS3. The LSODA method fails to capture all periodic oscillations in the true solution for both parameter sets. 

The RK23 method produces biologically unreasonable simulations for PS2. Moreover, the BDF and RK45 methods eventually become unbounded for PS3. This instability is the result of the carrying capacity becoming a threshold value when $r(t)$ becomes negative \cite{BoyceDiPrima2001}. In this case, the value $K$ becomes the minimum value for a population to persist unbounded instead of a maximum value. If the population is numerically estimated to be above $K$ and $r(t) < 0$, the population grows unbounded.
Neither of these issues are observed for the odeint, LSODA, or manual RK4 methods, and was likely unobserved before due to previous research considering only nonnegative values of $r(t)$ \cite{VanceCoddington1989, Banks1994, JiangShi2005}. The vast differences in qualitative and quantitative behavior raise concerns about the current accuracy of numerical methods used in built-in solvers for nonlinear, non-autonomous differential equations. 


\subsection{Periodic growth rate and carrying capacity}

The numerical issues observed from the built-in methods for simulating Eq.~(\ref{eq: model}) are further exacerbated when a time-varying carrying capacity is added. When a periodicaly-varying carrying capacity is added to Eq.~(\ref{eq: model}), then,

\begin{subequations}
\renewcommand{\theequation}{\theparentequation.\arabic{equation}}
    \label{eq: model 2}
    \begin{equation}
        P'(t) = r(t) P(t) \left(1 - \frac{P(t)}{K(t)}\right),
    \end{equation}
    \noindent where,
    \begin{equation}
         K(t) = K_b - K_s \cos\left(\frac{2 \pi t}{365} \right),   
    \end{equation}
\end{subequations}

\noindent and $r(t)$ as before. The $K_b$ component represents the baseline carrying capacity, and $K_s$ represents the carrying capacity's amplitude scaling factor. To avoid singularities, it is assumed that $0 < K_s < K_b$.  Few have explored the case where both parameters varying in time \cite{NisbetGurney1976, GrozdanovskiShepherdStacey2009}, but it is of interest to understand how external influences affect both components of the model.

A maximum cutoff value for $P(t)$ is implemented in the manual RK4 method to avoid potentially unbounded behavior that may occur with negative net growth rates. The selected cutoff value is the supremum of the carrying capacity, $\sup K(t) = K_b + K_s$, which denotes the maximum biologically relevant carrying capacity. Such biological constraints are easy to implement in manually-programmed numerical solvers, but difficult to incorporate for built-in solvers that have a black box framework. It can pay to write a numerical solver specifically designed to respect known theoretical features \cite{MilnerPugliese1999}.

\begin{figure}[h]\centering
  \includegraphics[width=\linewidth]{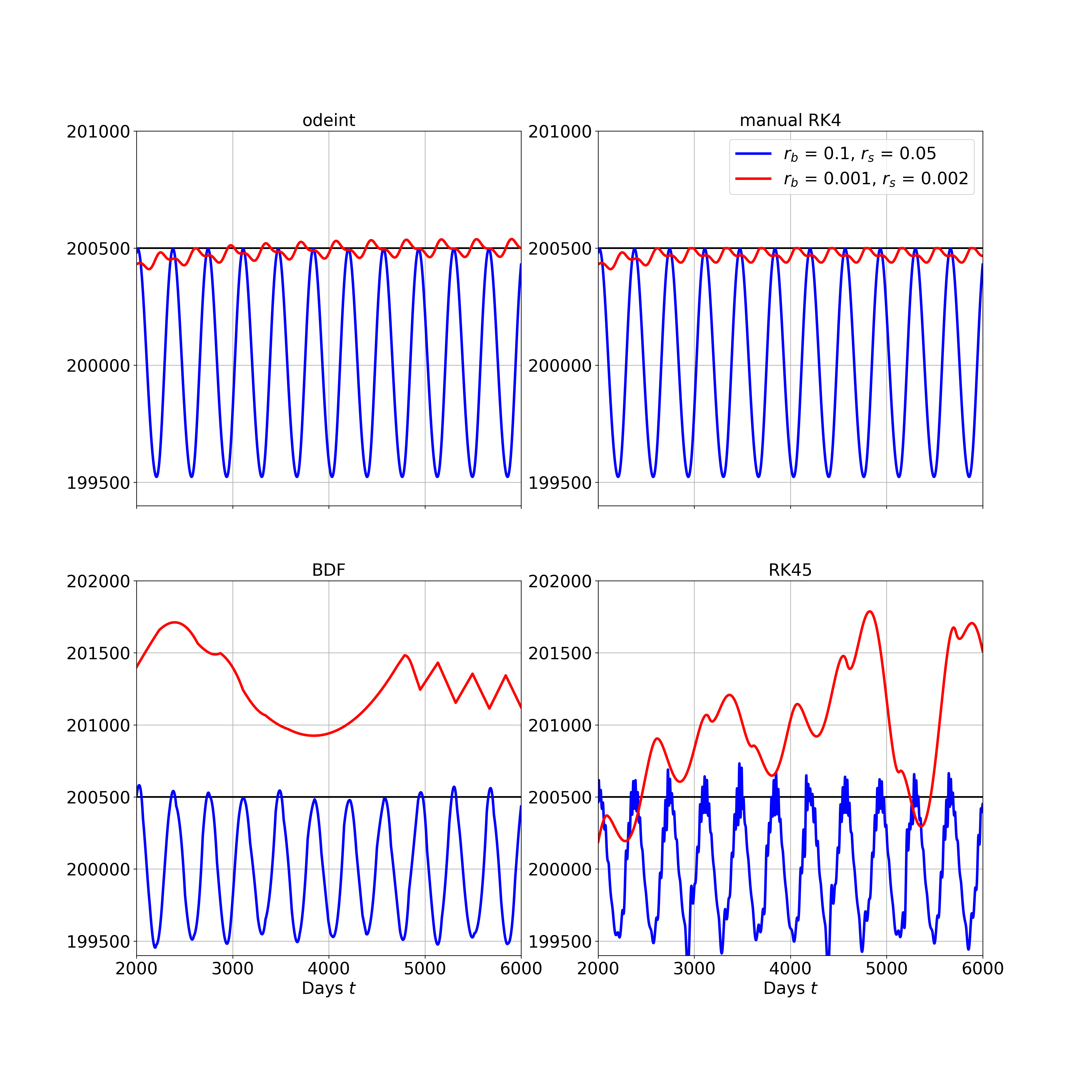}
  \caption{Simulations of Eq.~(\ref{eq: model 2}) using the odeint, manual RK4, BDF, and RK45 numerical solvers. Blue lines denote simulations with $r_b = 0.1$ and $r_s = 0.05$. Red lines denote simulations with $r_b = 0.001$ and $r_s = 0.002$. All simulations have $K_b = 200,000$, $K_s = 500$, and initial condition $P(0) = 200,000$. The solid black lines denote the maximum theoretical carrying capacity, $\sup K(t) = K_b + K_s$.}
  \label{fig: compare both}
\end{figure}

Although an explicit analytic solution does not exist for Eq.~(\ref{eq: model 2}), intuition expects the simulations to produce oscillatory behavior. Simulations of Eq.~(\ref{eq: model 2}) using the odeint, RK45, BDF, and manual RK4 methods are shown for two parameter sets in Figure \ref{fig: compare both}. Results from the odeint and manual RK4 methods produce consistent oscillations for both parameter sets, while the RK45 and BDF methods produce erratic or non-smooth behavior. The BDF and RK45 tests estimate values above $\sup K(t)$, even when the net growth rate remains nonnegative.

When comparing results between the odeint and manual RK4 methods, the time-varying growth rate does not affect the oscillations in $K(t)$ when $\inf r(t) > 0$, which aligns with the conclusions found in \cite{NisbetGurney1976}. Both tests also show $\sup P(t) < \sup K(t)$.  When $r(t)$ switches signum and the oscillatory components of $r(t)$ and $K(t)$ are \textit{in-phase}, the odeint simulation produces values exceeding $\sup K(t)$. This issue is not observed for the manual RK4 method due to the implemented cutoff for $P(t)$. Overall, the manually-programmed RK4 method can provide biologically reasonable behavior for a wider range of parameter values than the odeint method.

\section{Discussion \& conclusions}

Numerous disciplines use logistic models of varying complexity, and it can be tempting to implicitly trust the outputs from standardized, built-in numerical solvers. Although standardized, built-in numerical solvers can accurately simulate a variety of differential equation problems while providing a user-friendly interface \cite{Hindmarsh1983}, they may not be robust for solving all non-autonomous problems. Python's built-in ODE solvers fail to capture the true solution of a non-autonomous logistic model with a periodically-varying net growth rate changing signum, meanwhile a manually-programmed fourth-order Runge-Kutta method provides a closer approximation to the true solution. Numerical issues are further compounded with the added complexity of employing a periodic carrying capacity.

The observed numerical issues from built-in solvers may be attributed to their black box nature. Firstly, the particular algorithms used for step size adjustment in the built-in methods appear to be sensitive to Eq.~(\ref{eq: model}) when the net growth rate is negative, as seen by the BDF, RK45, and RK23 methods producing unbounded results for some Parameter Sets. All of the built-in methods tested allow for step size adjustment, whereas the manual RK4 method uses a fixed step size and outperformed the built-in solvers for a range of growth rate parameters. Secondly,  the built-in methods were not flexible to the same range of parameter values for the more complex non-autonomous model Eq.~(\ref{eq: model 2}). By incorporating the supremum of the carrying capacity to be the maximum value of the population in the manually-programmed method, simulations can produce biologically reasonable results when $r(t)$ and $K(t)$ are both \textit{in-phase} and \textit{out-of-phase}. 

The black box framework of built-in numerical solvers makes it difficult to both identify the exact source of numerical issues as well as incorporate biological constraints. On the other hand, it is easier to diagnose numerical issues in manually-programmed solvers because the user knows how the inputs are being manipulated within the solver. Although results from this example only capture a specific case of non-autonomous logistic growth, it shows that even the most frequently used and trusted built-in numerical solvers may not be robust for some nonlinear and non-autonomous systems. Moreover, this example suggests that careful consideration is warranted when simulating nonlinear, non-autonomous differential equations that do not have an analytical solution for comparison, particularly for models that have direct physical or biological meaning. The aforementioned issues will be useful to consider when improving existing standardized numerical solvers.

As the availability of data expands, the application of non-autonomous models becomes increasingly relevant for mechanistically modeling data-driven processes. The numerical issues encountered for this simple example of incorporating a periodic growth rate in a logistic model may foreshadow similar issues when modeling other applications which exhibit harmonic motion \cite{Morin2008} or may not have an analytical solution for comparison. The awareness of numerical solver accuracy becomes paramount to the validation and interpretation of mathematical models.

\section*{Acknowledgement}
This work was supported by the U.S. Department of Energy through the Los Alamos National Laboratory. Los Alamos National Laboratory is operated by Triad National Security, LLC, for the National Nuclear Security Administration of U.S. Department of Energy (Contract No. 89233218CNA000001). Research presented in this article was supported by the Laboratory Directed Research and Development program of Los Alamos National Laboratory under project number 20210062DR.

\end{document}